\documentclass[twocolumn,twoside,pra]{revtex4}

\usepackage{graphicx}
   \graphicspath{{./img/}}
\usepackage{amsmath}
\usepackage{amssymb}
\usepackage{bm}
\usepackage{units}
\usepackage{wasysym}

\newcommand{\im}{i}	%\mathrm{i}}
\newcommand{\diff}{d}	%\mathrm{d}}
\newcommand{\e}{e}	%\mathrm{e}}
\newcommand{\version}{February 13, 2008}
\DeclareMathOperator{\trace}{tr}

\begin{document}
%==========================================================================
\title{Fidelity and Entanglement of a Spatially Extended Linear Three-Qubit
Register}
%==========================================================================

\author{Roland Doll$^\ast$, Martijn Wubs$^\mathsection$,
Sigmund Kohler$^\ast$, and Peter H\"anggi$^\ast$}

\address{%
${}^\ast$Institut f{\"u}r Physik, Universit{\"a}t Augsburg,\\
Universit\"atsstra{\ss}e 1, D-86135 Augsburg, Germany}
\address{%
${}^\mathsection$
The Niels Bohr Institute, Niels Bohr International Academy \& QUANTOP, \\
Blegdamsvej 17, DK-2100 Copenhagen, Denmark}

\begin{abstract}
We study decoherence of a three-qubit array coupled to substrate
phonons.  Assuming an initial three-qubit entangled state that would
be decoherence-free for identical qubit positions, allows us to focus
on non-Markovian effects of the inevitable spatial qubit separation.
It turns out that the coherence is most affected when the qubits are
regularly spaced. Moreover, we find that up to a constant scaling
factor, two-qubit entanglement is not influenced by the presence of
the third qubit, even though all qubits interact via the phonon field.
\end{abstract}

\date{\version}

\maketitle

% ----------------------------------------------------------------------------
\section{Introduction}

A major obstable on the way towards a working quantum computer is
decoherence: the interaction of the qubits with their environment
reduces the indispensable quantum coherence of the quantum states.
Several strategies are pursued to beat decoherence.
An active strategy is quantum error correction, which requires a
redundant encoding of a logical qubit by several physical
qubits~\cite{Shor1995a,Kempe2005a,Stean2006a}.
Standard error correction protocols presuppose that all physical
qubits couple individually to uncorrelated baths. A passive strategy is the
use of decoherence-free subspaces (DFS)~\cite{Palma1996a, Zanardi1997a,
Lidar1998a, Lidar2003a}.
There, one logical qubit is encoded by several physical qubits in
such a way that the logical qubit states do not couple to the
environment.  Ideal DFSs occur when physical qubits couple via a
collective coordinate to a common bath.

For solid-state qubits, the coupling to
substrate phonons often is a relevant source of decoherence,
in particular for charge qubits in quantum dots \cite{Hayashi2003a}.
Whether these qubits experience correlated or uncorrelated noise
depends on their distance in relation to the coherence length of
the phonons, the sound velocity, the cutoff frequency and
also on the dimensionality of the substrate.
In Ref.~\cite{Doll2006a}, this dependence has been worked out by studying pure
dephasing of a two-qubit state with an initial entanglement that is
decoherence-free if both qubits couple to the environment at the same position
\cite{Yu2002a}. If the qubits are spatially separated, however, this behavior
changes: the entanglement decays until the transit time of a sound wave from one
qubit to the other is reached. However, if the qubits are embedded in a
quasi-one-dimensional environment, the noise at the two positions
eventually becomes sufficiently correlated to bring decoherence to a 
standstill. In this way, a decoherence-poor
subspace can emerge. Similar results can be found for the decoherence of
entangled states of a regularly spaced qubit array \cite{Doll2007b}.

This work is motivated by two questions: First, do imperfections in the
regular alignment of the qubits involve additional decoherence, or, put
differently, how do irregularities in the spatial extension of a qubit register
influence the collective decoherence properties? And second, is the entanglement
of a qubit pair affected by its indirect interaction with a third qubit via the
environment?

%------------------------------------------------------------------------------
\section{Qubits coupled to a bosonic field}\label{secModel}

%-------------
\begin{figure}
\begin{center}
\includegraphics{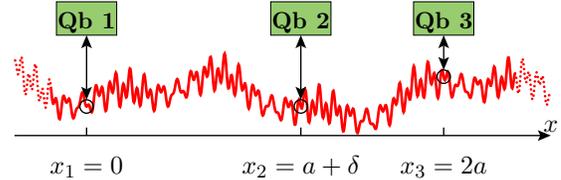}
\end{center}
\caption{\label{figCouplingToString}
Sketch of three qubits (green boxes) in a linear arrangement at
positions $x_\nu$, $\nu = 1,2,3$, with the distances $a+\delta$ and
$a-\delta$. They interact via a coupling to the bosonic field (red line).}
\end{figure}%
%-------------

%
\begin{figure*}
\includegraphics{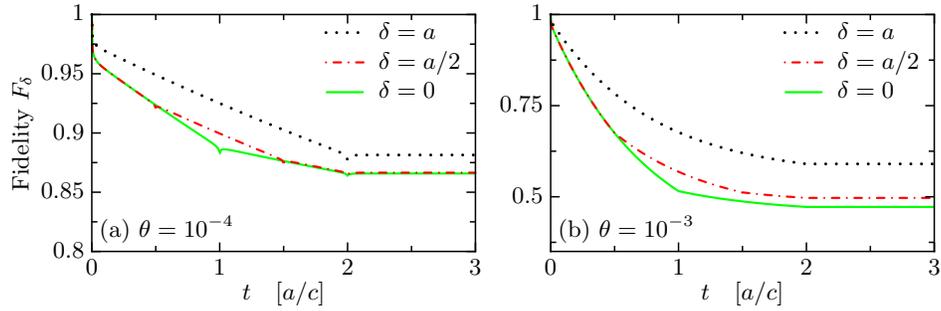}
\caption{\label{figFidelityVardist} Time evolution of the fidelity
\eqref{Frobust} for the temperatures $\theta = k_\text B T / \hbar \omega_\text
c = 10^{-4}$ (a) and $\theta = 10^{-3}$ (b), the qubit-field
coupling strength $\alpha = 0.001$, and various displacements
$\delta$ of the middle qubit (see Fig.~\ref{figCouplingToString}).}
\end{figure*}

As sketched in Fig.~\ref{figCouplingToString}, we consider a linear arrangement
of three qubits at positions $x_1=0$, $x_2=a+\delta$, and $x_3=2a$,
i.e.\ the nearest-neighbor separations $x_{12}=a+\delta$ and $x_{23}=a-\delta$.
To elaborate on the impact of spatially correlated noise we assume the qubit
array to be embedded in a channel-like structure, as may be realized in carbon
nanotubes or in linear ion traps. Thus, we treat the bosonic
environment as effectively one-dimensional. The total Hamiltonian
modelling
this situation reads
$H = H_\mathrm{q} + H_\mathrm{qb} + H_\mathrm{b}$,
where $H_\mathrm{q}= \sum_{\nu = 1}^3 \hbar\Omega_\nu \sigma_{\nu z}/2$
describes three qubits with energy splittings $\hbar \Omega_\nu$ with
$\sigma_{\nu z}$ being a Pauli matrix for qubit $\nu = 1,2,3$. Note that there
is no direct interaction between the qubits. The bosonic field described by
$H_\mathrm{b} = \sum_k \hbar \omega_k b_k^\dagger b_k$ consists of modes $k$
with energies $\hbar\omega_k$ and the respective annihilation and creation
operators $b_k$ and $b_k^\dagger$. We assume a linear dispersion relation
$\omega_k= c|k|$ with sound velocity $c$. The transit time of a
field distortion between the qubits $\nu$ and $\nu'$ is then $t_{\nu\nu'} =
x_{\nu\nu'}/c$ with $x_{\nu\nu'} = |x_\nu - x_{\nu'}|$.
Qubit $\nu$ couples linearly via the operator $X_\nu$ to
the field, so that the coupling Hamiltonian reads
\begin{equation} \label{eqnCoupling}
H_\mathrm{qb} =  \hbar \sum_{\nu = 1}^2 X_\nu \xi_\nu\,,
\end{equation}
with
$\xi_\nu = \xi(x_\nu) = \sum_k g_k \e^{\im k x_\nu} (b_k +b_{-k}^\dagger)$
the bosonic field operator at the respective qubit position $x_\nu$. We assume
the microscopic coupling constants $g_k$ to be real-valued, isotropic, and identical
for all qubits, i.e.\ $g_{k\nu} = g_k$ and $g_{-k} = g_k$. They determine the
spectral properties of the bath and show up in the spectral
density $J(\omega) = \sum_k g_k^2 \delta(\omega - ck)$. Here we consider an Ohmic
spectral density
\cite{Leggett1987a}
\begin{equation}\label{ohmic}
J(\omega) = \alpha\, \omega e^{-\omega/\omega_\text{c}} ,
\end{equation}
where the dimensionless parameter $\alpha$ denotes the overall coupling strength
and $\omega_\text{c}$ the cutoff frequency of the bath spectrum.

The dynamics for the total density operator $R$ of the qubits plus the
environment is governed by the Liouville-von Neumann equation
\begin{equation} \label{eqnLiouvillevonNeumann}
\im\hbar\frac{\diff}{\diff t} \widetilde R(t)
= \bigl[\widetilde H_\text{qb}(t), \widetilde R(t)\bigr]\,.
\end{equation}
The tilde denotes the interaction-picture representation with respect
to $H_0 = H_\mathrm{q}+H_\mathrm{b}$,
i.e.\ $\tilde A(t) = U_0^\dagger(t) A\, U_0(t)$, with time-evolution operator  $U_0(t) = \exp\{-i
H_0 t/\hbar\}$. We assume that at time $t=0$, the qubits can be
prepared in a well-defined initial state, uncorrelated with the
thermal bath. This constitutes an initial condition of the Feynman-Vernon type,
where the total initial density matrix
$\tilde R(0)$ is a direct product of a qubit and bath density operator,
$ \tilde R(0) = \tilde \rho(0) \otimes\rho^\mathrm{eq}_\mathrm{b}\,.$
The canonical ensemble of the bosons at temperature $T$ is denoted by
$\rho^\mathrm{eq}_\mathrm{b} = \exp(-H_\mathrm{b}/k_\mathrm{B}T)/Z$,
with $Z$ the partition function. We are interested in
the reduced density matrix of the qubits $\tilde\rho(t) =
\trace_\mathrm{b} \tilde R(t)$, where $\trace_\mathrm{b}$ denotes the
trace over the bath variables.

\section{Dephasing and entanglement decay}

In order to exemplify the impact of a spatial qubit separation on
decoherence, we consider as the initial state the three-qubit entangled $W$ state
  \begin{equation}\label{eqnWState}
  |W\rangle =
  \frac{1}{\sqrt{3}}\bigl(\, |100 \rangle + |010\rangle +|001\rangle\,\bigr)\,,
  \end{equation}
i.e.\ $\tilde \rho(0) = | W \rangle\langle W |$,
with the computational basis $\{|n_1n_2 n_3 \rangle\}$ where
$\sigma_{\nu z} |n_1 n_2 n_3\rangle = (-1)^{n_\nu} |n_1n_2n_3\rangle$ and
$n_\nu = 0,1$. Our motivation to focus on the initial
state \eqref{eqnWState} is twofold: First, $W$ states play an important
role in several protocols for quantum information
processing \cite{Joo2003a,Han2002a,Agrawal2006a}, so that the
entanglement dynamics after their preparation
is relevant in itself. Second,  the $W$ state is special since it stays robust under collective
dephasing, i.e.\ for vanishing qubit separations ($x_{1}=x_{2}=x_{3}$) \cite{Yu2002a,Doll2007b}.

Pure phase noise will also be assumed  in the following, in which case
the coupling operators in Eq.~\eqref{eqnCoupling} become $X_\nu
=\sigma_{\nu z}$. As a consequence, the interaction-picture
qubit operators remain time-independent, $\widetilde X_\nu(t) = X_\nu$.
The exact time evolution of the reduced
density operator can then be obtained, e.g.~by a direct solution of the
Liouville-von Neumann equation \eqref{eqnLiouvillevonNeumann} \cite{Doll2007b}.
Amazingly, the exact result can even be obtained with an \textit{approximative}
time-local master equation approach already  in second order of the qubit-field
coupling $\alpha$ \cite{Doll2007c}. It turns out that the
density matrix elements in the basis $\{|n_1n_2 n_3 \rangle\}$ at time $t$
are proportional to their initial values. Thus, all density matrix
elements that are initially zero remain zero, so that for the state
$| W \rangle$, the dissipative quantum dynamics is restricted
to the states
\begin{equation}
\label{eqnMatrixElementOneQubitExited}
|1\rangle = |1 0 0 \rangle,\quad
|2\rangle = |0 1 0 \rangle,\quad
|3\rangle = |0 0 1 \rangle\,,
\end{equation}
i.e.~at most nine out of $64$ density matrix elements are nonvanishing.
Initially they are all equal, i.e.~$ \rho_{jj'}(0) = \langle j|
\rho(0)|j'\rangle=1/3$, with $j,j'=1,2,3$. They evolve as
\begin{equation}\label{rho_dynamics}
\tilde\rho_{jj'}(t) = \frac{1}{3}
\exp\{-\Lambda_{jj'}(t) + \im [\phi_j(t) - \phi_{j'}(t)]\}\,,
\end{equation}
where the real part $\Lambda_{jj'}(t)$ of the exponent accounts for
the time-dependent amplitude damping of the matrix element; indirect
interactions among qubits via the environment give rise to a
time-dependent frequency shift and the concomitant phase
$\phi_j(t)-\phi_{j'}(t)$ with $\phi_j(t) \equiv \varphi_j(t) -
\Delta\Omega_j t$.
Here $\Delta\Omega_j$ is a static frequency renormalization $\Omega_j \to
\Omega_j + \Delta\Omega_j$ for qubit $j$ and $\varphi_j(t)$ describes
its onset; cf.\ Ref.~\cite{Solenov2007a}.
Henceforth we work in the interaction picture with respect to the renormalized
energies. With the scaled temperature $\theta = k_\text B T / \hbar
\omega_\text c$ and the scaled times $\tau = \omega_\text c t$ and
$\tau_{\nu\nu'} = \omega_\text c t_{\nu\nu'}$, the density matrix
elements become $\tilde\rho_{jj'}(\tau) = \exp\{-\Lambda_{jj'}(\tau) + \im
[\varphi_j(\tau )-\varphi_{j'}(\tau)]\}/3$ \cite{Doll2007b}
with the phases
\begin{align}
\varphi_j(\tau) &= -\frac{\alpha}{2} \sum_{\nu,\nu'=1}^3
(-1)^{\delta_{j\nu}+\delta_{j\nu'}} \sum_\pm \arctan[\tau\pm\tau_{\nu\nu'}]\,,
\end{align}
and the amplitude damping $\exp\{-\Lambda_{jj'}(\tau)\} =
f(\tau,\tau_{jj'})/f(\tau,0)$, where
\begin{align}
f(\tau,\tau') &=
\frac{\left|\Gamma(\theta[1-\im\tau'])\right|^{16\alpha}}{\left|\Gamma^2(\theta[1+\im(\tau-\tau')])
\Gamma^2(\theta[1+\im(\tau-\tau')])
\right|^{4\alpha}}\\ \nonumber
&\quad \times \frac{1}{|1+\tau^{2}[1-\im\tau']^{-2}|^{4\alpha}}\,.
\end{align}
%
%%%%%%%%%%%%%%
%%%%%%%%%%%%%%

As expected for pure dephasing, populations are preserved:
$\tilde\rho_{jj}(t) = \tilde\rho_{jj}(0)$.  This implies that
neither the qubits nor the total system will
reach thermal equilibrium.  
Nevertheless decoherence does occur since relative phases between
eigenstates will be randomized so that off-diagonal density matrix
elements decay.

\begin{figure}
\begin{center}
\includegraphics{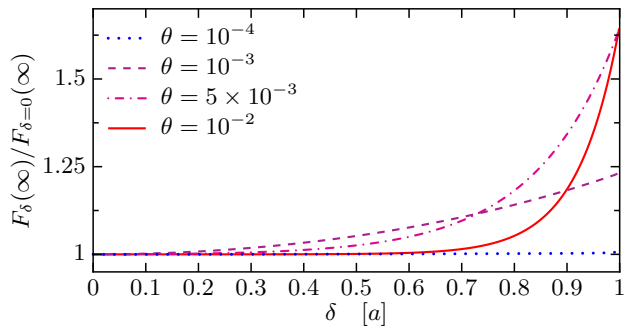}
\end{center}
\caption{\label{figSaturatedFidelityVardist}Final fidelity
$F_\delta(\infty)$ as a function of the displacement $\delta$ of the
middle qubit, scaled to the value for equidistant qubit
arrangement $\delta = 0$ for various
temperatures $\theta = k_\text B T / \hbar \omega_\text c$ and coupling
strength $\alpha = 0.001$.}
\end{figure}

Only in the absence of the environment ($\alpha = 0$), the qubits remain in
the $W$ state~\eqref{eqnWState}. As a measure for the
deviation from this ``ideal'' output state $\rho(0)=|W\rangle\langle
W|$, we employ the fidelity $F(t) =\trace\{ \tilde\rho(t) \tilde
\rho(0)\}$ \cite{Poyatos1997a}, which in our case reads
\begin{equation}\label{Frobust}
\begin{split}
{F}_\delta(t) &= \frac{1}{3}\sum_{j,j^\prime=1}^{3} \tilde
\rho_{jj^\prime}(t) \\
&=\frac{1}{3} + \frac{2}{9}
\sum_{j<j'}\e^{-\Lambda_{jj'}(t)} \cos[\varphi_j(t) - \varphi_{j'}(t)] \,.
\end{split}
\end{equation}
The index $\delta$ refers to the displacement of the middle
qubit $\nu = 2$.
The time evolution of the fidelity is shown in
Figs.~\ref{figFidelityVardist}a,b for two different temperatures.
Clearly, the fidelity decay is slowed down whenever a transit time $t=t_{jj'}$
is reached. At time $t=t_{13}$, when the field has also enabled
communication between the two outer qubits $1$ and $3$, decoherence
even comes to a standstill! Note that other initial states may lead to
complete dephasing \cite{Solenov2007a}.
The fidelity saturates to a finite value $F_\delta(t> t_{13}) =
F_\delta(\infty)$ which is larger the lower the temperature.  For a fixed
temperature, the stable fidelity increases if the middle qubit is
displaced away from the central position and is maximal for  $\delta =
\pm a$, i.e.\ if qubit $2$ becomes co-located with qubit 1 or 3.
The fidelity gain as a function of the asymmetry $\delta/a$ is shown
in Fig.~\ref{figSaturatedFidelityVardist}.

The explanation of this behavior of $F_\delta(\infty)$ follows from
Eq.~\eqref{Frobust}. The individual coherences $\tilde \rho_{\nu \nu'}$
decay approximately exponentially and with the same decay rate~\cite{Doll2007b},
but stop decaying at different times $|x_{\nu}-x_{\nu'}|/c$. 
Hence for the fidelity it pays off to displace the middle qubit,
thereby stopping the decay of $\tilde\rho_{23}$ at an earlier time
$(a-\delta)/c$ at the expense of stopping the decay of
$\tilde\rho_{12}$ only at time $(a+\delta)/c$.

For the extreme cases
$\delta=0$ and $\delta=a$, the fidelity decreases monotonously with
temperature, but for
$0 < \delta < a$ an optimal temperature exists for which the fidelity gain is maximal. This behavior is related to the fact that for all temperatures $\tilde \rho_{12}(t)=\tilde\rho_{23}(t)$ when $\delta=0$, and $\tilde \rho_{13}(t)=\tilde\rho_{23}(t)$ when $\delta=a$. Only in the general asymmetric configuration  do all three coherences stop decaying at different times.
%%%%%%%%%%%%%%%%%
%%%%%%%%%%%%%%%%

The initial $W$ state is special in the sense that it exhibits
bipartite entanglement between any qubit pair. But does the  middle qubit  affect the
entanglement decay of the outer ones in any way?
 The reduced density matrix of qubits $1$ and $3$ is
\begin{equation}
\begin{split}
\trace_2 \tilde\rho(t) &= \frac{1}{3}\left( |00\rangle\langle00| +
|01\rangle\langle01| + |10\rangle\langle10| \right) \\
&\quad + \tilde\rho_{13}(t) |10\rangle\langle01| + \tilde\rho_{31}(t) |01\rangle\langle10|\,.
\end{split}
\end{equation}
It depends on time only through the coherence $\tilde\rho_{13}(t) =
\tilde\rho^\ast_{31}(t)$ and thus, according to
Eq.~\eqref{rho_dynamics}, only on the transit time $t_{13} = 2a/c$
between the outer qubits. The entanglement of the outer qubits is
therefore \textit{independent} of the displacement $\delta$ of the middle
qubit! Indeed, for their concurrence \cite{Wootters1998a} we find $C[\trace_2
\tilde\rho(t)] = 2 |\tilde \rho_{13}(t)|$. The same dynamics is found
for the two-qubit concurrence when only the
outer two qubits had been present, with the two-qubit $W$ state (i.e.
the robust Bell state) as their initial state~\cite{Doll2006a}.
Surprisingly, the only effect of the presence of the middle
qubit is a rescaling of the  concurrence of the outer two by a factor
$2/3$.

\section{Conclusions}
We have studied the pure dephasing of three spatially separated qubits
in an Ohmic environment.  Dephasing can be incomplete when starting in
a $W$ state, not only for symmetrically  spaced qubits.  For fixed
separation of the outer two qubits, the final fidelity is even larger
the more  ``clustered'' the qubits are.
%, a conclusion that will hold for $N$ qubits as well. 
Surprisingly, the middle qubit does not affect the dynamics of the
concurrence of the outer two.

\begin{acknowledgments}
This work has been supported by DFG via SFB 631.  PH and SK
acknowledge funding by the DFG excellence cluster ``Nanosystems
Initiative Munich''.
\end{acknowledgments}

%------------------------------------------------------------------------------

\end{document}